\documentclass[12pt]{article}
\usepackage{graphicx}
\usepackage{amsmath,amssymb,amsfonts}
\usepackage{slashed}

\begin{document}

\title{The Riemann-Silberstein vector in the Dirac algebra}

\author{Shahen Hacyan
}

\renewcommand{\theequation}{\arabic{section}.\arabic{equation}}

\maketitle
\begin{center}

{\it  Instituto de F\'{\i}sica,} {\it Universidad Nacional Aut\'onoma de M\'exico,}

{\it A. P. 20-364, Cd. de M\'exico, 01000, Mexico.}

\end{center}
\vskip0.5cm

\begin{abstract}

It is shown that the Riemann-Silberstein vector, defined as ${\bf E} + i{\bf B}$, appears naturally in the
$SL(2,C)$ algebraic representation of the electromagnetic field. Accordingly, a compact form of the Maxwell
equations is obtained in terms of Dirac matrices, in combination with the null-tetrad formulation  of general
relativity. The formalism is fully covariant; an explicit form of the covariant derivatives is presented in terms
of the Fock coefficients.

\end{abstract}


key words: Maxwell equations, Dirac matrices algebra

\section{Introduction}

The Riemann-Silberstein (RS) vector is defined as the complex sum of the electric and magnetic field vectors:
${\bf E}+i{\bf B}$. It appeared in 1907 in an article by Silberstein \cite{silb}, and was applied many years later
by various authors to different problems \cite{bbbb,bibi,rusos} (see, in particular, Ref. \cite{bibi} for a
historical account and full bibliography).

The RS vector appears conspicuously in electromagnetism because it describes the electromagnetic field in a
particular representation of the Lorentz group, namely an irreducible representation of the $SL(2,C)$ group. This
is evident if the method of spin coefficients is applied to the Maxwell equations.

As for the applications of spinor algebra, several formulation of the Maxwell and Einstein equations have been
proposed following the pioneering article of Newman and Penrose \cite{np}. A particulary compact formulation was
worked out by Plebanski \cite{pleb} in the seventies, based on the use of a null tetrad as a system of reference
(see Ernst \cite{ernst} for its relation with other authors formulations).

The aim of the present article is to further elucidate the  role of the RS vector in the context of spinorial
calculus. For this purpose, the null tetrad formalism of general relativity is used in combination with Dirac
spinors, \textit{i.e.} four-components spinors, and the related  matrices of the Dirac algebra. The RS can thus be
identified as the spinorial image of the electromagnetic field in this particular representation. Being fully
covariant, our approach is valid in any Riemannian space-time. Furthermore, it generalizes to the Maxwell
equations a previous work on the Dirac equation in curved space-time \cite{yo}. The result is a particularly
compact and covariant form of the Maxwell equations that can be used in combination with the Dirac equation in
problems of general relativity.

\section{Maxwell equations and Dirac matrices}

The Dirac matrices $\gamma^{\alpha}$ are such that
\begin{equation}
\gamma^{\alpha}\gamma^{\beta}+\gamma^{\beta}\gamma^{\alpha}=-2 g^{\alpha \beta} ,
\end{equation}
where $g^{\alpha \beta}$ is the metric tensor (signature $\{-+++\}$ and $c=1$, in the following)

In the chiral gauge, for instance, they take the form
\begin{equation}
\gamma^0  = \left(
                \begin{array}{cc}
                  \mathbf{0} & -I \\
                  -I & \mathbf{0} \\
                \end{array}
              \right) \label{14},
\end{equation}
\begin{equation}
\gamma^i  = \left(
                \begin{array}{cc}
                  \mathbf{0} & \sigma^i \\
                  -\sigma^i & \mathbf{0} \\
                \end{array}
              \right), \label{15}
\end{equation}
where $\sigma^i$ are the usual Pauli matrices.

Let $A_{\mu}$ be the electromagnetic potential and $f_{\alpha \beta}= \partial_{\beta} A_{\alpha}-
\partial_{\alpha} A_{\beta}$ the electromagnetic tensor. In flat space and Cartesian coordinates ($f_{01}=E_x, ~f_{12}=-B_z$, etc.), we have
\begin{equation}
\gamma^{\mu} A_{\mu} \equiv \slashed{A}= \left(
                \begin{array}{cc}
                  \mathbf{0} & A^0+\boldsymbol{\sigma} \cdot {\bf A}  \\
                 A^0- \boldsymbol{\sigma} \cdot {\bf A} & \mathbf{0} \\
                \end{array}
              \right) \label{17}
\end{equation}
and it follows from the definition of ${\bf E}= -\frac{\partial}{\partial t} {\bf A}-\nabla A^0$ and ${\bf
B}=\nabla \times {\bf A}$ in terms of $A^{\mu}$ --which are equivalent to the homogeneous  Maxwell equations--:
\begin{equation}
-\slashed{\partial} \slashed{A}  = \left(
                \begin{array}{cc}
                  \boldsymbol{\sigma} \cdot ({\bf E} +i{\bf B})& \mathbf{0}  \\
                 \mathbf{0} & -\boldsymbol{\sigma} \cdot ({\bf E} -i{\bf B}) \\
                \end{array}
              \right) \equiv \mathbb{F}, \label{193}
\end{equation}
if the Lorentz gauge,
$$
\nabla_{\mu} A^{\mu} = \frac{\partial}{\partial t} A^0 + \nabla \cdot {\bf A} =0,
$$
is used.

Accordingly, the inhomogeneous Maxwell equations take the compact form
\begin{equation}
- \slashed{\partial} \mathbb{F}= \slashed{\partial} \slashed{\partial} \slashed{A} = 4\pi \slashed{J},
\end{equation}
where  $J^{\mu} = (\rho , {\bf J})$ is the electromagnetic current

We thus see that the RS vector, defined as
\begin{equation}
{\bf F} \equiv {\bf E} +i{\bf B},
\end{equation}
appears naturally in the above representation of the electromagnetic field.

Defining
$$\sigma^{\alpha \beta}=\frac{1}{2} (\gamma^{\alpha}\gamma^{\beta}-\gamma^{\beta}\gamma^{\alpha}),$$
it simply follows that
\begin{equation}
\frac{1}{2} f_{\mu \nu} \sigma^{\mu \nu} = \mathbb{F} =  \left(
                \begin{array}{cc}
                  \boldsymbol{\sigma} \cdot {\bf F}& \mathbf{0}  \\
                 \mathbf{0} & -\boldsymbol{\sigma} \cdot {\bf F}^* \\
                \end{array}
              \right).
\end{equation}
The components of $f_{\mu \nu}$ will be identified in the context of the null-tetrad formalism (see below).

\subsection{The RS vector}

Defining the invariants of the field:
$$
{\bf E}^2 - {\bf B}^2 \equiv{\cal E}^2 - {\cal B}^2, \quad  {\bf E} \cdot {\bf B} \equiv{\cal E} {\cal B},
$$
we have
$
{\bf F}^2 = ({\cal E}+ i {\cal B} )^2,
$
and it follows that
\begin{equation}
\mathbb{F}^2  =  \left(
                \begin{array}{cc}
                  ({\cal E}+ i{\cal B})^2 I & \mathbf{0}  \\
                 \mathbf{0} & ({\cal E}- i{\cal B})^2 I \\
                \end{array}
              \right).
\end{equation}
Thus $\mathbb{F}^2$ is totally diagonal.

In the particular case of a null-electromagnetic field, ${\cal E}=0={\cal B}$, the matrix $\mathbb{F}$ turns out
to be nilpotent of degree 2: $\mathbb{F}^2=0$.

In the general case, the matrix $ \mathbb{F}$ satisfies the equation
\begin{equation}
\mathbb{F}^4-2 ({\cal E}^2 - {\cal B}^2)\mathbb{F}^2 + ({\cal E}^2 + {\cal B}^2)^2 \mathbb{I}=0,
\end{equation}
implying that the eigenvalues $\lambda$ of $\mathbb{F}$ are
$$
\lambda^2 = ({\cal E}\pm i{\cal B})^2.
$$
Accordingly we have in general 4 eigenvalues, $\lambda_{(i)} = \pm ({\cal E}\pm i{\cal B})$, with 4 eigenfunctions
$\psi_{(i)}$ such that
$$
\mathbb{F} \psi_{(i)} = \lambda_{(i)} \psi_{(i)} ~\quad i=(1...4).
$$
Furthermore, since $\mathbb{F}^2$ is completely diagonal, its eigenvectors can be taken as any set of four
linearly independent spinors $u_{(i)}$, namely
$$
\mathbb{F}^2 u_{(i)} = \lambda_{(i)}^2  u_{(i)}. $$ It then follows that, in general,
\begin{equation}%
\psi_{(i)}= (\mathbb{F}+\lambda_{(i)})u_{(i)} ,\label{eisp}
\end{equation}
which can be interpreted as a generalization to Dirac spinors of the two-components Bloch spinors (if the
$u_{(i)}$ are chosen as constant units spinors).

\section{Null tetrad formalism}

The null-tetrad is  a set of null-vectors $e^a_{\alpha}$ defining one-forms $e^a= e^a_{\mu}~dx^{\mu}$, such that
$e^1$ and $e^2$ are complex conjugates to each other, $e^3$ and $e^4$ are real, and
$$
ds^2 = g_{\alpha \beta} ~dx^{\alpha}~dx^{\beta} = \eta_{ab}~ e^a e^b,
$$
where
$$
\eta_{ab}=\eta^{ab} =\left(
                       \begin{array}{cccc}
                         0 & 1 & 0 & 0 \\
                         1 & 0 & 0 & 0 \\
                         0 & 0 & 0 & 1 \\
                         0 & 0 & 1 & 0 \\
                       \end{array}
                     \right)
$$

As shown in Ref. \cite{yo}, a convenient choice of the Dirac matrices in the null-tetrad formalism is
$$
\gamma^1 = \sqrt{2} \left(
  \begin{array}{cccc}
    0 & 0 & 0 & 0 \\
    0 & 0 & 1 & 0 \\
    0 & 0 & 0 & 0 \\
    -1 & 0 & 0 & 0 \\
  \end{array}
\right) \quad \gamma^2 = \sqrt{2} \left(
  \begin{array}{cccc}
    0 & 0 & 0 & 1 \\
    0 & 0 & 0 & 0 \\
    0 & -1 & 0 & 0 \\
    0 & 0 & 0 & 0 \\
  \end{array}
\right)
$$
\begin{equation}
\gamma^3 = \sqrt{2} \left(
  \begin{array}{cccc}
    0 & 0 & 0 & 0 \\
    0 & 0 & 0 & -1 \\
    -1 & 0 & 0 & 0 \\
    0 & 0 & 0 & 0 \\
  \end{array}
\right) \quad \gamma^4 = \sqrt{2} \left(
  \begin{array}{cccc}
    0 & 0 & 1 & 0 \\
    0 & 0 & 0 & 0 \\
    0 & 0 & 0 & 0 \\
    0 & 1 & 0 & 0 \\
  \end{array}
\right),\label{D}
\end{equation}
satisfying the condition
$$
\gamma^a\gamma^b + \gamma^b\gamma^a= -2 \eta^{ab} \mathbb{I}.
$$

With the above choice of Dirac matrices,  it follows that in (standard) Cartesian coordinates
\begin{equation}
\slashed{\partial} =  \gamma^n \partial_n = \sqrt{2}\left(
  \begin{array}{cccc}
    0 & 0 & \partial_4 & \partial_2 \\
    0 & 0 & \partial_1 & -\partial_3 \\
    -\partial_3 & -\partial_2 & 0 & 0 \\
    -\partial_1 & \partial_4 & 0 & 0 \\
  \end{array}
\right),
\end{equation}
where the directional derivatives $\partial_n$ are
$$
\partial_1 = \frac{1}{\sqrt{2}} (\partial_x +i \partial_y) , \quad \partial_2 = \frac{1}{\sqrt{2}} (\partial_x -i
\partial_y),
$$
\begin{equation}
\partial_3 = \frac{1}{\sqrt{2}} (\partial_z + \partial_t) , \quad \partial_4 = \frac{1}{\sqrt{2}} (\partial_z-
\partial_t).
\end{equation}

The associated matrices $\sigma^{ab}= \frac{1}{2}(\gamma^a\gamma^b - \gamma^b\gamma^a)$ were given in \cite{yo};
here we repeat them in the appendix for the sake of completeness. From their explicit form, it follows that for
the electromagnetic tensor $f_{ab}$, in particular, and for \emph{any} antisymmetric tensor, $f_{ab}=-f_{ba}$, in
general,
\begin{equation}
f_{ab}\sigma^{ab}=2\left(
  \begin{array}{cccc}
    f_{12}+f_{34} & -2f_{42} & 0 & 0 \\
    2f_{31} & -f_{12}-f_{34} & 0 & 0 \\
    0 & 0 & f_{12}-f_{34} & -2f_{32} \\
    0 & 0 & 2f_{41} & -f_{12}+f_{34} \\ \label{matriz}
  \end{array}
\right).
\end{equation}

Comparing with \eqref{193}, we see that the cartesian components of the RS vector ${\bf F}$ are
$$
F_x+iF_y= 2 f_{31} ,\quad F_x-iF_y= -2 f_{42}, \quad F_z = f_{12}+f_{34},
$$
\begin{equation}
F^*_x+iF^*_y= -2 f_{41} ,\quad F^*_x-iF^*_y= 2 f_{32}, \quad F^*_z = -f_{12}+f_{34}.
\end{equation}

\subsection{General coordinates system}

In a general system of coordinates, the directional derivatives $\partial_n$ must be replaced by the covariant
directional derivative $\nabla_n$.  The Dirac equation in a general coordinates system can thus be written as
\begin{equation}
\gamma^n \nabla_n \psi + i m \psi =0
\end{equation}
in terms of the covariant derivative
\begin{equation}
\nabla_n= \partial_n \psi + \Gamma_n,
\end{equation}
where $\Gamma_n$ are the Fock coefficients \cite{fock,chap}. In a tetradial representation, they are defined as
$$
\Gamma_n = - \frac{1}{4} \Gamma_{abn}\sigma^{ab},
$$
where $\Gamma^a_{~bc}$ are the Ricci rotation coefficients given by $$de^a= e^b \wedge \Gamma^a_{~b},$$ with
$\Gamma^a_{~b} = \Gamma^a_{~bc} e^c$, and $\Gamma_{abc}= -\Gamma_{bac}$ (see, e.g. Refs. \cite{pleb,ernst}). In
the null-tetrad formalism, their forms follow from \eqref{matriz} (see also \cite{yo})
.

For a rank two tensor, in particular, we have
\begin{equation}
\nabla_c f_{ab} = \partial_c f_{ab} - \Gamma^n_{~ac} f_{nb} -\Gamma^n_{~bc} f_{an}.\label{Gamma}
\end{equation}
Applying this formula to our matrix $\mathbb{F}$, one finds after some lengthy but straightforward algebra that
\begin{equation}
\nabla_n \mathbb{F}=  \partial_n \mathbb{F} +{\bf \Gamma}_n ~\mathbb{F}-\mathbb{F}~  {\bf \Gamma}_n~,
\end{equation}
a formula that can also be checked by direct substitution.

Accordingly, the inhomogeneous Maxwell equations take the form
\begin{equation}
\gamma^n \nabla_n \mathbb{F}=  4\pi {\slashed J},
\end{equation}
valid in general.

\section{Concluding remark}

The above analysis clarifies the role of the Riemann-Silberstein vector in the context of a spinorial approach to
classical electromagnetism. Given the full covariance of all the formulas obtained in this paper, the present
formulation can be applied in future publications to problems in general relativy involving electromagnetism and
Dirac fields.

\section*{Appendix}

The  matrices $\sigma^{ab}$ associated to the Dirac matrices \eqref{D} are \cite{yo}
$$
\sigma^{12}+\sigma^{34} =2 \left(
  \begin{array}{cc}
    \begin{array}{cc}
             1 & 0 \\
             0 & -1\\
           \end{array} & \mathbf{0} \\
    \mathbf{0} & \mathbf{0}
              \\
  \end{array}
\right)
                           \quad
\sigma^{31} =2 \left(
  \begin{array}{cc}
    \begin{array}{cc}
             0 & 0 \\
             1 & 0\\
           \end{array} & \mathbf{0} \\
    \mathbf{0} & \mathbf{0}
              \\
  \end{array}
\right)
\quad \sigma^{42} =2 \left(
  \begin{array}{cc}
    \begin{array}{cc}
             0 & -1 \\
             0 & 0\\
           \end{array} & \mathbf{0} \\
    \mathbf{0} & \mathbf{0}
              \\
  \end{array}
\right)
$$
\begin{equation}
-\sigma^{12}+\sigma^{34} =2 \left(
  \begin{array}{cc}
    \mathbf{0} & \mathbf{0} \\
   \mathbf{0} & \begin{array}{cc}
             -1 & 0 \\
             0 & 1 \\
           \end{array}
              \\
  \end{array}
\right)
                           \quad
\sigma^{32} =2 \left(
  \begin{array}{cc}
    \mathbf{0} & \mathbf{0} \\
    \mathbf{0} & \begin{array}{cc}
             0 & -1 \\
             0 & 0 \\
           \end{array}
              \\
  \end{array}
\right)
\quad \sigma^{41} =2 \left(
  \begin{array}{cc}
    \mathbf{0} & \mathbf{0} \\
    \mathbf{0} & \begin{array}{cc}
             0 & 0 \\
             1 & 0 \\
           \end{array}
              \\
  \end{array}
\right).
\end{equation}

\end{document}